\documentclass[12pt]{iopart}
\input{epsf.tex}
\eqnobysec
\begin{document}
\begin{flushright}ULB-TH/02-36\\DCPT-02/85\\{\tt hep-th/0212085}
\end{flushright} 
\title{Fermionic Subspaces of the Bosonic String}

\author{Auttakit Chattaraputi\dag
\footnote[5]{Email: aut@deathstar.phys.sc.chula.ac.th} Fran\c cois   
Englert\ddag
\footnote[6]{Email: fenglert@ulb.ac.be}  
 Laurent Houart\S \footnote[7]{Research Associate F.N.R.S., Email:
lhouart@ulb.ac.be} and Anne Taormina$\|$\footnote[8]{Email:
anne.taormina@durham.ac.uk} }  

\address{\dag\  Department of Physics, University of Chulalongkorn Bangkok
10330, Thailand}

\address{\ddag\ Service de Physique Th\'eorique, Universit\'e Libre de
Bruxelles,
 Campus Plaine, C.P.225, Boulevard du Triomphe, B-1050 Bruxelles, Belgium}

\address{\S\  Service de Physique Th\'eorique et Math\'ematique,  
Universit\'e Libre de Bruxelles, Campus Plaine C.P. 231, Boulevard du
Triomphe, B-1050 Bruxelles, Belgium}

\address{$\|$ Department of Mathematical Sciences, University of Durham,
 South Road, DH1 3LE Durham, England }

\begin{abstract} A universal symmetric truncation of the bosonic string
Hilbert space yields all known closed fermionic string theories in ten
dimensions, their D-branes and their open descendants. We highlight the
crucial role played by group theory and two-dimensional conformal field
theory in the construction and emphasize  the predictive power of the
truncation. Such circumstantial evidence points towards the existence of a
mechanism which generates space-time fermions out of bosons dynamically
within the framework of bosonic string theory. \footnote[9]{Presented at the
Leuven RTN-Workshop, Belgium, September 2002 and at the International
Conference on Conformal Field Theory, Chernogolovka, Russia, September 2002.
}

\end{abstract}

\maketitle
\setcounter{equation}{0}
\section{Introduction and conclusion}

It is well-known that ten-dimensional fermionic strings can be analyzed in
terms  of  bosonic operators, a consequence of the boson-fermion
equivalence in two dimensions. The approach taken here is different. We
show that the Hilbert space of the  bosonic string compactified on
suitable sixteen dimensional tori contain subspaces with fermionic degrees
of freedom. This programme was initiated in 1986 in the framework of
closed strings
\cite{cent}. We revisited the approach in the last two years
\cite{eht01,ceht02} and extended the construction to the open string
sectors. The recent developments in the Conformal Field Theory 
description of open strings
\cite{sagr} are instrumental to our results.

  We determine the fermionic subspaces by performing a truncation  of the
bosonic Hilbert space.  To ensure consistency of the truncation for open
string sectors (and hence for  D-branes) we impose a {\em symmetric}
truncation in both closed string sectors. This consistency condition lifts
all ambiguities about the fermionic subspaces found by truncation and, most
importantly, allows the approach to be {\em predictive}. We can classify
all the fermionic subspaces and  exhibit how this classification is
related to global properties of the group $SO(16)$. All fermionic strings
live in subspaces of the bosonic string compactified on sublattices of the
$E_8 \times SO(16)$ weight lattice and the sublattice $E_8 \times E_8$ of 
$E_8 \times SO(16)$ contains the supersymmetric theories IIA and IIB as
well as the heterotic superstrings. All  non-heterotic strings and all
their (stable and unstable) D-branes are classified by the discrete
subgroups of the centre of
$\widetilde{SO}(16)$. Significantly, the characteristic properties of
fermionic D-branes  (tension, charge conjugation, chirality changing
T-dualities) are predicted from purely bosonic considerations.
Furthermore, the Chan-Paton  groups of tadpole-free open  fermionic
strings are also correctly  obtained via truncation, and in particular,
the anomalies in type I  do cancel.

 Truncation provides a dictionary translating all fermionic string
properties to bosonic string ones. If a non-perturbative mechanism exists
which isolates the fermionic subspaces, the scope of the M-theory quest
would be considerably enlarged: there would be no elementary fermions at a
fundamental level and supersymmetry would have a dynamical origin.

\section{Symmetric truncation} The truncation of the bosonic string Hilbert
space which yields all its ten-dimensional fermionic subspaces is highly
constrained. A first constraint originates in the closed string sector,
where coherence of the theory imposes that modular invariance be
preserved by truncation, while a second constraint emerges from the open
string sector, where one must require the truncation to be consistent with
boundary conditions relating the left and right moving  closed strings, i.e.
with the introduction of D-branes. The resulting truncation must be {\em
symmetric} and we now review how it works.

 We perform a toroidal compactification of the 26-dimensional closed
bosonic string theory at an enhanced symmetry point with gauge group
${\cal G}_L \times {\cal G}_R$, with 
${\cal G}_L, {\cal G}_R$ two  semi-simple, simply laced Lie groups. With
both groups of  rank $d=24-s$, $0 \le s \le 24$, the compactified bosonic
theory lives in
$s+2$ dimensions, and the original transverse Lorentz group $SO(24)_{tr}$ 
becomes the Lorentz group $SO(s)_{tr}$, which does not possess the
spinorial representations needed to accommodate space-time fermions, and 
which cannot therefore play the role of the transverse Lorentz group of a
fermionic theory in
$s+2$ dimensions. In order to manufacture an appropriate Lorentz group, one
uses a stringy analog of the field theoretical  construction which turns
isospin into spin in four dimensional gauge theory \cite{thooft}. Namely,
one requires that
${\cal G}_L$ and ${\cal G}_R$ (in the heterotic case only ${\cal G}_R$)
admit an
$SO(s)_{int}$ subgroup, and one takes as new transverse Lorentz group the
diagonal
$SO(s)_{diag}$ group with algebra
\begin{equation}\label{diagonal} so(s)_{diag}={\rm diag}~ [so(s)_{tr}
\times so(s)_{int}]
\end{equation}  generated by $J^{ij}=L^{ij}+K_0^{ij}, i < j, i,j =
1,...,s$.  The algebraic set-up Eq.(\ref{diagonal}) is a first step in
creating  spin  from isospin. The second  step is to ensure the
closure of the full Lorentz algebra in $s+2$ dimensions. This can be done
only if all states corresponding to 12 compact dimensions are removed,
except for some zero-modes, and the maximum value of $s$ accommodating
fermions turns out to be 8. Although these facts follow from the highly
non-trivial closure condition of the Lorentz algebra, they can be
understood in simpler qualitative terms. Indeed, the existence of
space-time fermions in a covariant formalism is rooted in the existence of
worldsheet supersymmetry. The central charge of the superghost is 11 and
 the timelike and longitudinal fermions contribute 1 to the central
charge. The hidden superconformal invariance in the light-cone gauge thus
requires the removal of 12 bosonic fields. The zero-modes kept in the 12
dimensions account for the oscillator zero-point energy which is equal to
$(-1/24) \times 12$ (in units $\alpha^\prime =1/2$) and which is to be
removed. Therefore zero-modes kept in 12 dimensions must contribute an
energy 1/2.  In this way, space-time fermions can be obtained provided a
truncation of the Hilbert space is performed. At this stage truncation is
done by hand, but the group theory classification of fermionic D-branes
and the host of correct predictions resulting from truncation strongly
suggest the existence of an underlying dynamics.

We now specify the truncation and restrict hereafter to $s=8$, i.e. to
10-dimensional fermionic strings.     The toroidal compactification
should therefore be performed   on the lattice of a Lie group of rank
$d=24-8=16$ with subgroup $SO(8)_{int}$.  The compactification lattice in
both sectors (or in the  right sector only for the heterotic strings) is
taken to be a sublattice of the $E_8 \times SO(16)$ weight lattice which
preserves the modular  invariance of the partition function, whose ${\cal
G}_L\times {\cal G}_R$ lattice contribution $P(\tau,\bar\tau)$ is
separately modular invariant and given by,

\begin{equation}
\label{factor} P(\tau,\bar \tau)=\sum_{\alpha,\beta} N_{\alpha\beta} \
\bar
\alpha_L(\bar\tau)\ \beta_R(\tau)\, ,
\end{equation} where 
\begin{equation}
\label{partition}
\beta_R (\tau)=\sum_{\sqrt{2\alpha^\prime}{\bi p}_{oR}\in \, (o)} \exp
\{2\pi i\tau[ {({\bi p}_{o R}+{\bi p}_{\beta R} )^2\over2} + N_R^{(c)}
-{\delta
\over 24 }]\}\, .
\end{equation} Here $\beta$ is a partition function for a sublattice
$(\beta)$ of the ${\cal G}_R= E_8 \times SO(16)$ weight lattice (i.e.
$(\beta)=(o)_{E_8} \oplus (i)_{16}, \: i=o,v,s,c$) and
${\bi p}_{\beta R}$ is a fixed vector, arbitrarily chosen, of the
sublattice
$(\beta)$.
$ N_R^{(c)}$ is the oscillator number in the $\delta=16$ compact
dimensions. A similar expression holds for $\bar
\alpha_L(\bar\tau)$,  $\bar\alpha$ labeling a partition function for a
sublattice of the weight lattice of ${\cal G}_L$. The coefficients
$N_{\alpha\beta}$ are 0 or 1  and are chosen in such a way that 
$P(\tau,\bar \tau)$ is modular invariant.

 In order to proceed with the truncation (exemplified here in the right
sector of the theory), we decompose the
$SO(16)$ factor of ${\cal G}_R$ in $SO^{\, \prime}(8)
\times SO(8)$ and first truncate all states created by oscillators in the
12 dimensions defined by the $E_8\times SO^{\,\prime}(8)$ root lattice. 
The group
$SO(8)$ is identified with the internal symmetry group $SO(8)_{int}$. As
discussed above, the closure of the new Lorentz algebra dictates we keep
zero-modes in the 16 compact dimensions in such a way that
\begin{equation}
\label{ghost} {1\over2}\, {\bi p_R}^2[E_8 \times SO(16)] = {1\over2}\,
{\bi p_R}^2[SO(8)] +{1\over 2}\, ,
\end{equation}  with ${\bi p_R}({\cal G})$  a vector of the weight
lattice of the group 
${\cal G}$. The zero-mode contribution $1/2$ in Eq.(\ref{ghost}) comes from
$SO^{\, \prime}(8)$ as there are no vectors of norm squared one in $E_8$.
The only zero-mode contributions from $E_8 \times SO(8)'$ we keep are two
fixed
$SO(8)'$ 4-vectors
${\bi p^\prime_v}$ and 
${\bi p^\prime_s}$, so that we truncate the lattice partition functions
according to,
\begin{eqnarray}
\label{truncations} &&o_{16} \rightarrow v_8\, , \qquad v_{16}\rightarrow
o_8
\, ,\nonumber\\ &&s_{16} \rightarrow -s_8\, ,
\quad ~ c_{16}
\rightarrow -c_8 \, .
\end{eqnarray} It follows from the closure of the Lorentz algebra that
states belonging to 
$v_8$ or $o_8$ are bosons while those belonging to the spinor partition
functions
$s_8$ and
$c_8$ are space-time fermions. In accordance with the spin-statistic
theorem we  have  flipped the sign in the partition function of the
space-time spinor partition functions.

All heterotic strings were obtained, using Eq.(\ref{truncations}),
 in reference \cite{lls}. 
To obtain all fermionic D-branes in the non-heterotic theories, we must
truncate {\em both}  sectors of the modular invariant partition functions
Eq.(\ref{factor})  according to
Eq.(\ref{truncations})\footnote[1]{Hence the terminology
`symmetric truncation'.} \cite{ceht02}. As the $E_8$ lattice is Euclidean
even self-dual, we concentrate on the
$SO(16)$ weight lattice. Their are four even self-dual Lorentzian $SO(16)$
lattices. The corresponding modular invariant partition functions are (modulo
the  contribution from the $E_8$ lattice and from the non-compact
dimensions),
\begin{eqnarray}
\label{BOB} OB_b&=&\bar o_{16}\ o_{16}+ \bar v_{16}\ v_{16}+
\bar s_{16}\ s_{16}+\bar c_{16}\ c_{16}\, ,\\ \label{BOA} OA_b&=&\bar
o_{16}\ o_{16}+ \bar v_{16}\ v_{16}+
\bar s_{16}\ c_{16}+\bar c_{16}\ s_{16}\, ,\\
\label{B2B} IIB_b&=&\bar o_{16}\ o_{16}+ \bar s_{16}\ o_{16}+
\bar o_{16}\ s_{16}+\bar s_{16}\ s_{16}\, ,\\ \label{B2A} IIA_b&=&\bar
o_{16}\ o_{16}+ \bar c_{16}\ o_{16}+
\bar o_{16}\ s_{16}+\bar c_{16}\ s_{16}\, .
\end{eqnarray} They yield, after symmetric truncation, the four consistent
non-heterotic ten-dimensional fermionic string partition functions, namely,\\
\begin{eqnarray}
\label{OB} OB_b \rightarrow  \bar v_8\ v_8+ \bar o_8\ o_8+
\bar s_8\ s_8+\bar c_8\ c_8 &\equiv& OB\, ,\\ \label{OA} OA_b
\rightarrow
 \bar v_8\ v_8+ \bar o_8\ o_8+
\bar s_8\ c_8+\bar c_8\ s_8 &\equiv& OA\, ,\\ \label{2B} IIB_b \rightarrow
\bar v_8\ v_8 - \bar s_8\ v_8 -
\bar v_8\ s_8+\bar s_8\ s_8 &\equiv& IIB\, ,\\
\label{2A} IIA_b \rightarrow \bar v_8\ v_8 - \bar c_8\ v_8 -
\bar v_8\ s_8+\bar c_8\ s_8 &\equiv& IIA\, . \end{eqnarray}

\section{Fermionic D-branes and torus geometry} The properties of the
bosonic D9-branes pertaining to the four different  theories compactified
on $E_8 \times SO(16)$ lattices can be related to the geometry of the
configuration space torus characterizing  each compactification. These tori
are linked to each other through global properties of the universal
covering group
$\widetilde{SO}(16)$ as we shall now show.

 The amplitudes ${\cal A}_{tree}$   describing  the D9-branes in the  tree
channel are obtained from the torus partition functions
Eqs.(\ref{BOB})-(\ref{B2A}) by imposing  Dirichlet boundary conditions on
the compact space. In the tree channel, the latter consists in the
following relation between compactified momenta,
\begin{equation}
\label{dirc} {\bi p_L}- {\bi p_R}=0\, ,
\end{equation} as well as in a match between left and right oscillators. 
The amplitudes of elementary bosonic D9-branes are  given in Table 1, both
in the tree channel and in its S-dual loop channel.
 
\begin{table}
\caption{Bosonic D9-brane amplitudes.}
\begin{indented}
\item[]\begin{tabular}{||c||c|c||}
\hline &$2^5{\cal A}_{tree}$&${\cal A}_{loop}$\\
\hline
$OB_b$&$(1/2)\, (o_{16}+ v_{16}+ s_{16}+ c_{16})$&$o_{16}$\\
\hline
$OA_b$& $o_{16}+ v_{16}$&$o_{16}+ v_{16}$\\
\hline
$IIB_b$& $o_{16}+ s_{16}$&$o_{16}+ s_{16}$\\
\hline
$IIA_b$& $2\, o_{16}$&$o_{16}+ v_{16}+ s_{16}+ c_{16}$\\
\hline
\end{tabular}
\end{indented}
\end{table}

In order to identify the configuration space torus on which each theory is
defined, recall that in the  conformal $\sigma$-model description of 
these theories in presence of torsion $b_{ab}$, the left and  right
momenta are given by 
\begin{eqnarray} {\bi p_R} &=& [{1\over2} m_b + n^a (b_{ab} + g_{ab})]
{\bi e^b},
\ \
\nonumber \\ {\bi p_L} &=&  [{1\over2} m_b + n^a (b_{ab} - g_{ab})] {\bi
e^b},
\label{close}
\end{eqnarray} where $\{{\bi e^a}\}$ is the lattice-dual basis of the basis
$\{{\bi e_a}\}$ defining the configuration space torus
\begin{equation}
\label{ptorus} {\bi x} \equiv {\bi x} +2 \pi n^a {\bi e_a} \qquad n^a \in
{\cal Z}\ , 
\end{equation} and the   lattice metric is given by $g_{ab}= {\bi e_a}.
{\bi e_b}$ . The weight vectors $2{\bi e_a}$  generate  four
sublattices
 of the weight lattice of SO(16). They can be read off from the second
column in Table 1, as ${\cal A}_{loop}$ yields the winding lattice $n^a{\bi
e_a}$. The classification of bosonic D-branes  can be visualized by the
volume-preserving projection of configuration space tori in Figure 1,  and
provides, after truncation, the classification of the ten-dimensional 
fermionic subspaces.

\begin{figure}
\begin{center}
\hskip.5cm\epsfbox{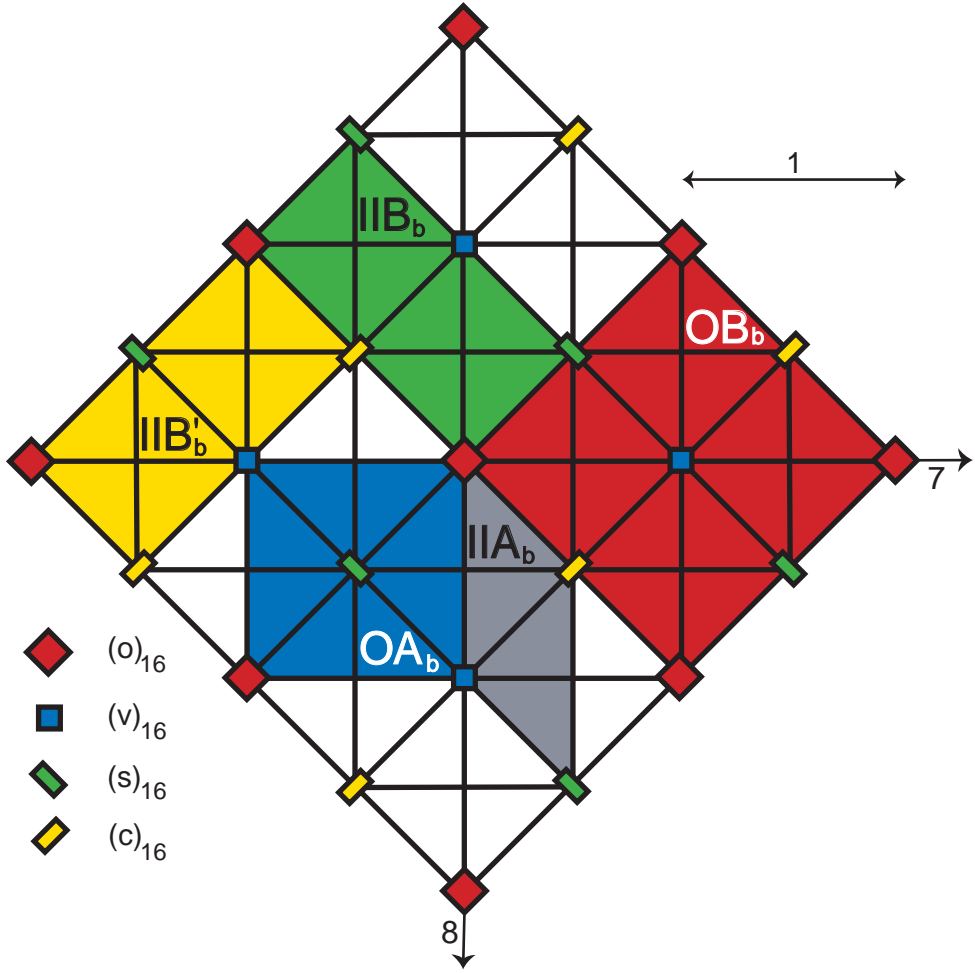}
\end{center}
\caption{\label{label}Projected weight lattice of $SO(16)$ in the plane of
the two orthogonal simple roots $r_7$ and $r_8$. The volumes
$\xi_{OB_b}=2,\xi_{OA_b}=\xi_{IIB_b}=1,\xi_{IIA_b}=\frac{1}{2}$ of the unit
cells, exhibited in shaded areas, must be multiplied by
$(2\pi)^8. 2^{-8}$ to yield the 
$SO(16)$ compactification space torus volume of the  four bosonic
theories  (in units where
$\alpha^\prime =1/2$). The theories IIB and IIB' are isomorphic and differ
by the interchange of $s_{16}$ and $c_{16}$}
\end{figure}

 The tori $\widetilde t$ of the four bosonic theories are, as group
spaces, the   maximal toroids $\widetilde{\cal T}/Z_c $ of the locally
isomorphic groups 
$E_8\times \widetilde{SO}(16)/Z_c$ where $Z_c$ is a subgroup of the centre 
$Z_2\times Z_2$ of the
 universal covering group
$\widetilde{SO}(16)$.  We write
\begin{eqnarray}
\label{coset}\widetilde t\, ( OB_b)& = &\widetilde{\cal T}\, ,\nonumber\\
\widetilde t\, (OA_b)& =&
\widetilde{\cal T}/Z^d_2\, , \nonumber\\ \widetilde t\, (IIB_b)& =
&\widetilde{\cal T}/Z^+_2 ~~~{\rm or}~~~
\widetilde{\cal T}/Z^-_2\, , \nonumber\\ \widetilde t\, (IIA_b)& =&
\widetilde{\cal T}/(Z_2 \times Z_2) \, , \end{eqnarray}  where
$ Z^d_2 = diag(Z_2 \times Z_2)$ and the superscripts $\pm$ label the two
isomorphic $IIB_b$ theories obtained by interchanging $(s)_{16}$ and
$(c)_{16}$.  There is thus a unified picture for the four theories related
to the global properties of the $SO(16)$ group.

\noindent {\bf {\em Tension of an elementary D9-brane}}: Since tree
amplitudes are proportional to the square of the D-brane tension $T$,
Table 1 provides the following relations between the tensions of the
elementary  D9-branes of the different theories
\begin{equation}
\label{trel}  \sqrt{2}\, T_{OB_b}=T_{OA_b}=T_{IIB_b}=(1/\sqrt{2})\,
T_{IIA_b}
\, .
\end{equation} To get their values, we recall that the tension 
$T^{bosonic}_{Dp}$ of a Dp-brane in the 26-dimensional uncompactified
theory is 
\cite{pob}
\begin{equation}
\label{dpb} T^{bosonic}_{Dp} = {\sqrt\pi\over2^4 
\kappa_{26}}(2\pi\alpha^\prime{}^{1/2})^{11-p}\, , \end{equation}
 where
$\kappa_{26}^2= 8\pi G_{26}$ and $G_{26}$ is the Newtonian constant in 26
dimensions. The tensions of the Dirichlet D9-branes of the four
compactified theories are obtained from Eq.(\ref{dpb}) by expressing
$\kappa_{26}$ in term of the 10-dimensional coupling constant
$\kappa_{10}$. Recalling that $\kappa_{26}=
\sqrt{V} \kappa_{10}$ where $V$ is the volume of the configuration space
torus, one finds,  using Figure 1,
\begin{eqnarray}
\label{tob} T^{}_{OB_b} = {\sqrt\pi\over \sqrt{2}
\kappa_{10}}(2\pi\alpha^\prime{}^{1/2})^{-6}\, ,\\ \label{toa} T^{}_{OA_b}
= T^{}_{IIB_b} ={\sqrt\pi\over \
\kappa_{10}}(2\pi\alpha^\prime{}^{1/2})^{-6}\, ,\\ \label{t2a}
T^{}_{IIA_b}={\sqrt{2}
\sqrt\pi\over
\kappa_{10}}(2\pi\alpha^\prime{}^{1/2})^{-6}\, . \end{eqnarray} These are
consistent with  Eq.(\ref{trel}).

 Truncation on the loop amplitudes of Table 1 yields the amplitudes
describing the fermionic D9-branes of respectively $OB, OA, IIB$ and $IIA$.
 Furthermore tension is conserved in the truncation as proven in
reference~\cite{eht01}. The tensions of the different bosonic D9-branes
given in Eqs.(\ref{tob})-(\ref{t2a}) are thus equal, when measured with
the same gravitational constant $\kappa_{10}$, to the  tensions of the
corresponding fermionic D9-branes \cite{pob,kets}. This is indeed a
correct prediction.

\noindent {\bf {\em Charge conjugation}}: Different D9-branes of a given
bosonic theory are joined by strings of minimal size ${\bi e_a}
=\frac{1}{2}{\bi w_a}$. Hence the number of distinct fermionic branes is
equal to the number of lattice points in the unit cell of the bosonic
torus. Charge conjugation for fermionic branes arises from a lattice shift
by
$(v)$ in the bosonic string. A glance at Figure 1 provides us with a
correct prediction of four charged stable fermionic D9-branes in OB (two
D-branes and their corresponding antibranes), two neutral unstable
D9-branes in OA, two charged stable D9-brane in IIB (one D-brane and its
antibrane) and one neutral unstable  D9-brane in IIA. 

\noindent {\bf {\em Fermionic Dp-branes}} ($p < 9$): Up to now, we have
considered D9-branes. The determination of fermionic D8-branes (and more
generally of lower dimensional even Dp-branes) by truncation appears at
first sight impossible because there seem to be  no bosonic
counterpart to the fermionic chirality changing T-duality which
relates fermionic D9-branes and D8-branes. The switching between $OA$
and $OB$ (or $IIA$ and $IIB$) theories by T-duality is imposed in fermionic
strings by worldsheet supersymmetry in the covariant formalism, or
equivalently by the closure of the Lorentz algebra in the light-cone
gauge. Hence truncation for even dimensional branes can  be consistent
with Lorentz invariance only if there exists an involution $\cal I$ in the
parent bosonic theories relating  $OA_b$ and $OB_b$ (or $IIA_b$ and
$IIB_b$) which would interchange the lattice partition functions of their
D-branes. Fermionic D8-branes would then be related by truncation, in
agreement with the requirement of Lorentz invariance, to  parent
bosonic branes obtained by submitting  bosonic D9-branes to  both $\cal
I$ and to a bosonic T-duality. Remarkably such an involution does exist as
we now explain.

In the description of toroidal compactification
of bosonic strings, E-duality maps a D-brane localized on
a torus onto a D-brane completely wrapped on it or vice versa~\cite{gpr}. At
an enhanced symmetry point E-duality does not necessarily map a given
lattice onto itself. Even more strikingly, starting from any Lagrangian
realization of $OA_b$ and $IIA_b$, E-duality {\em always} maps
$OA_b$ onto $OB_b$ and 
$IIA_b$ onto $IIB_b$ \cite{ceht02}. Such a duality we call {\em odd
E-duality} to distinguish it from even E-dualities which map a lattice
onto itself. Lagrangian realizations connected by odd E-dualities are
called odd realizations. While for $OA_b$ and $IIA_b$ there are only odd
realizations, for
$OB_b$ and $IIB_b$ one may have both even and odd realizations. These
differ by  inequivalent antisymmetric fields $b_{ab}$. Thus the required
involution is realized by the odd E-duality and the correct partition
functions of even fermionic Dp-branes are obtained, in accordance with
Lorentz invariance, from the truncation of the D(p+8)-brane resulting from
 combining an odd E-duality on the $SO(16)$ torus with a bosonic
T-duality.

We thus see that, as a consequence of the existence of odd
E-dualities, chirality changing T-dualities in fermionic strings are
encoded in the bosonic string !

\section{Tadpole-free open descendants}

Space-filling D-branes may be used to define  open string theories. These are
plagued by  massless tadpoles which give rise to divergences. In the
uncompactified 26-dimensional bosonic string, these divergences can be
eliminated by a restriction to unoriented strings and introducing a
Chan-Paton group
$SO(2^{13})$. Geometrically this amounts to take $2^{12}$ D25-branes
(+images) to cancel the negative tension of an $O25$ orientifold. We
now explain how  truncation yields all the tadpole-free fermionic open
descendants. 

Compactification of the unoriented bosonic string at an enhanced symmetry
point generically reduces the rank of the Chan-Paton group ensuring tadpole
cancellation. In addition, symmetry breaking may occur because D9-branes
may sit at different locations in the lattice.  The explicit computation
of the Chan-Paton multiplicities is done by requiring the cancellation of
the tadpoles arising in the tree channel of the annulus, the M\"obius
strip and the Klein bottle. Furthermore, one proves that the Chan-Paton
group is conserved by truncation. We obtain in this way the Chan-Paton
groups listed in Table~2 \cite{eht01,ceht02}.

It is particularly remarkable that for type I where the tadpole would induce
a genuine anomaly, the correct Chan-Paton group follows from bosonic
considerations only.

\begin{table}
\caption{Open descendants, $N$ is the number of D-branes (+images).}
\begin{indented}
\item[]\begin{tabular}{||c||c|c||}
\hline &Chan-Paton group&$N$\\
\hline
$OB_b\to OB \to B$ &$[SO(32-n)\times SO(n)]^2$&$2^6$\\
\hline
$OA_b\to OA \to A$& $SO(32-n)\times SO(n)$&$2^5$\\
\hline
$IIB_b\to IIB \to I$& $SO(32)$&$2^5$\\
\hline
\end{tabular}
\end{indented}
\end{table}

\ack 
This work was supported in part  by the NATO grant PST.CLG.979008.

\end{document}